# Planting trees at the right places: Recommending suitable sites for growing trees using algorithm fusion

**Pushpendra Rana[1] and Lav R. Varshney[2]**

[1]Indian Forest Service, Shimla, India
Affiliated Researcher, University of
Illinois, Urbana Champaign
pranaifs27@gmail.com

[2]University of Illinois at Urbana-
Champaign, United States
varshney@illinois.edu

## Abstract

Large-scale planting of trees has been proposed as a low-cost natural solution for carbon mitigation, but is hampered by poor selection of plantation sites, especially in developing countries. To aid in site selection, we develop the ePSA (e-Plantation Site Assistant) recommendation system based on algorithm fusion that combines physics-based/traditional forestry science knowledge with machine learning. ePSA assists forest range officers by identifying blank patches inside forest areas and ranking each such patch based on their tree growth potential. Experiments, user studies, and deployment results characterize the utility of the recommender system in shaping the long-term success of tree plantations as a nature climate solution for carbon mitigation in northern India and beyond.

1. ## Introduction

The recent push to promote large-scale tree planting as a climate change mitigation solution has fueled a fierce debate around the nature and intensity of social-economic and ecological impacts of these programs (Fleischman et al., 2020; Malkamäki et al., 2018). Supporters favor massive tree-planting due to its cost-effectiveness and global carbon sequestration potential, whereas opponents focus on adverse effects on local livelihoods, biodiversity, and ecology (Bastin et al., 2019; Busch et al., 2019; F. Fleischman et al., 2020; Veldman et al., 2019). A middle ground, however, advocates tree planting only in places where it meets local community needs, suits local environs, and restores/enriches local forest ecosystems (Brancalion & Holl, 2020; Holl & Brancalion, 2020).

Machine learning (ML)-based models can assist forest rangers in choosing socially and ecologically appropriate sites for planting trees similar to ML applications in other fields such as prioritizing animal feed locations (Handan-Nader & Ho, 2019) and prioritizing environmental inspections (Hino et al., 2018). ML can use large social, spatial, and ecological datasets to learn patterns and then identify areas appropriate for growing trees. Moreover, ML can enable systems to continuously update their decisions based on new data from decisions and responses of users (Appel et al., 2014; Salganik, 2019; Varshney, 2016). We tried to use ML classifiers to assist tree planting decisions in northern India but found it quite challenging for several reasons.

First, plantation data is highly skewed and imbalanced. Official data only covers sites selected for planting trees (positive labels) and we usually lack information about sites that rangers have rejected for planting (negative labels). Second, historical datasets on plantations do not have spatial information and so we do not know exactly where planting (sites of positive labels) happened. Moreover, plantation records are not systematic and uniform across the landscape, leading to potential biases. Third, there is significant heterogeneity in the size of plantation areas from one place to another so universal predictive plantation model is not appropriate for all sites. Fourth, forest officials may game the system to serve their vested interests by omitting certain plantation areas as possibilities; this lack of accountability and transparency can severely hurt plantation recommendation engines. Finally, there is a lack of large-scale, fine-resolution data on factors that influence site selection. The several sources of spatial data from government or international platforms are inadequate due to poor resolution and incompatibility among several geospatial layers. Moreover, limited financial resources and manpower restrict the large-scale collection of soil survey data and other biophysical factors at a finer scale.

Remote sensing data coupled with GIS has previously been used to locate potential tree-planting sites (Cuong et al., 2019; Wu et al., 2008), evaluate suitability of land for tree plantations (Basir, 2014) and agroforestry (Ahmad et al., 2017), and specify tree species under tree plantations (Y. Chen et al., 2019; Lahssini et al., 2015). ML has also been used to evaluate land suitability for agriculture and forestry (Sarmadian et al., 2014; Loi, 2008), to predict site index (Sabatia & Burkhart, 2014), and to predict future land use transition areas and scenarios (Grinand et al., 2020).

Notwithstanding these previous papers, purely data-driven systems are limited (Hofman et al., 2017; Selbst et al., 2019) due to social-ecological complexity and large-scale heterogeneity that limit gathering of real or (realistic) training data that covers the universe of possibilities in a forestry landscape. If a range of factors, interactions, and feedbacks that affect planting decisions are not captured properly in data-driven models and the complex forest system is abstracted too much, the resulting incorrect classifications will have critical safety impacts on livelihoods, biodiversity, and carbon storage (Hofman et al., 2017; Mueller et al., 2019; Selbst et al., 2019; Thompson et al., 2012).

Planting site suitability classifiers are only loosely supported by data for areas of the feature space unobserved in training data, leading to wild extrapolation in such areas (Kshetry & Varshney, 2019). Even methods such as covariate shift, domain adaptation, meta-recognition (Scheirer et al., 2011), reject options or estimating the confidence of deep learning models (Mallick et al., 2020) fail to contribute much in the presence of large-scale epistemic data uncertainty, as here. This points to not formulating planting site selection as a pure ML problem.

Instead, here we take a physics-based AI approach for an e-Plantation Site Assistant (ePSA) that combines forestry science knowledge with an ML classifier in a form of algorithm fusion to control epistemic uncertainty and maintain AI decision safety (FAO, 1984; Hofman et al., 2017; Kshetry & Varshney, 2019; Selbst et al., 2019). The idea is to combine theoretical and expert forestry knowledge with learned models to cover not only the known factors that affect planting decisions but also, the other known or unknown factors that affect these decisions but for which we do not have data for purely ML models. Specifically, we use traditional forestry knowledge to guide and weight features that affect site suitability scores for growing trees and use the ML-model to capture large landscape-level deforestation dynamics. The overall AI system explores available blank areas for growing trees and the potential for effective tree growth and survival due to congenial biophysical, social, and edaphic factors.

We choose Himachal Pradesh, one of the northern states in India for the research. Government of India has incentivized the mountainous states of northern India (and other states with substantial forest cover) in the form of green bonus or extra funds to protect forest cover and support tree planting through various programs and schemes (Busch & Mukherjee, 2018). The aim is to motivate these northern states to protect and improve their forests through tree planting so that their forestry landscape, which form an important water catchment drainage zones for northern agricultural belt of India, remain protected. ePSA, therefore, can potentially help Government of India to effectively implement and monitor the progress of tree planting in water catchment zones to protect soil erosion and to boost water flow to agricultural regions of northern India for increasing agricultural productivity.

Since 2002, Himachal Pradesh has spent an estimated 248.24 million US dollars on tree-planting programs, covering an area of 236,686 Ha (*Himachal Forest Statistics*, 2019). We have chosen Himachal Pradesh to test ePSA recommendation engine for two main reasons. First, Himachal Pradesh is one of the most developed and well-governed states in northern India and is ranked number two after Kerala in many human development indicators. All the requisite conditions for developing an effective ePSA recommendation engine are in place in Himachal Pradesh. If ePSA fails in the relatively favorable social, economic, and political contexts of Himachal Pradesh, this system will likely be ineffective in more challenging contexts of the other northern Indian states and other parts of India or many similar developing country contexts. Second, the state of Himachal Pradesh showed its commitment to move forward with this application and provided government spatial and social datasets and expert guidance to develop ePSA to prioritize tree planting patches across the state as per their tree growth potential.

In our recommendation engine, we use traditional forestry knowledge to consider existing land uses by local communities, places where natural constraints restrict tree growth, and areas where existing dense vegetation limits tree planting (FAO, 1984). We also include rules based on the extent and change of tree cover, slope, aspect, elevation, soil quality, and nearness to habitations in our scoring algorithm to predict site suitability of each studied forest grid. These rules originate from theory, past scholarship, and technical studies (FAO, 1984) (Supplementary Material). We used satellite data from various national and international sources, performed feature engineering, and created relevant indicators that may influence the site suitability of the location. The entire process is explained in the Methods section and can potentially be a useful guide for other researchers to meaningfully use remote sensing data in their studies. We have used QGIS, ArcGIS, and Python capabilities to process the geospatial data and to make it usable for developing this application. We employed ML trained on a large set of features ($n = 31$) to capture the landscape-level deforestation dynamics to properly account for factors that lead to tree cover loss in any landscape, a process that is largely outside the ability of human-derived traditional knowledge to capture.

To build the final recommendation system, we aggregate with 90% weight for traditional forestry knowledge and 10% weight for ML. We give only small weight to ML due to epistemic uncertainty and limited training data. We rely heavily on theory and expert rule-based systems, since they are easy to implement and their decisions are easily interpreted/explained to users in the form of domain-specific rules. Moreover, since factors that drive site suitability and availability of land change very slowly,

theory- and expert- systems capture long-term tacit knowledge that new data-driven models cannot. Informal hyperparameter tuning also shows this 90%-10% combination works well (for details, please refer to Section 2.4). More ML training data and formal hyperparameter tuning may be used to improve the AI system.

Our recommendation engine is deployed as a mobile app and is now being tested on the ground to evaluate the site selection choices of field rangers. Experiments, user surveys, and experience from this deployment show both the potential and the limitations of this technology, especially in promoting accountability and transparency on the ground. The use of our recommendation engine has huge policy implications and may inform the direction of tree planting in India and beyond. It may save money and resources by assisting forest rangers to select the best tree planting sites for carbon mitigation.

## 2. Methods

We used an algorithm fusion methodology (Kshetry & Varshney, 2019) to create the plantation site recommendation engine as in Figure 1, with a further two-stage process. The first stage leverages rule-based and machine-learning algorithms with remote sensing data to predict site suitability, and the second stage uses those predictions to prioritize the most appropriate tree planting sites as per site suitability values. Our methodology involves the following steps.

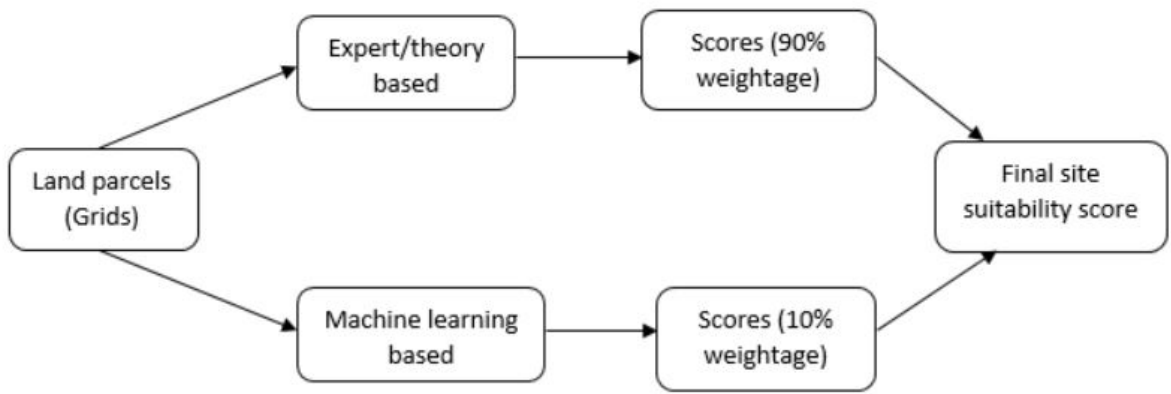

Fig. 1. Schematic diagram of the algorithm fusion approach to incorporate both expert/theory-based and machine learning in calculating the final suitability score for each forest grid.

### *2.1 Creation of grids (forest patches)*

We created 265 meters square grids (7.0225 hectares) for the entire state of Himachal Pradesh using QGIS for recommending growing of trees at the grid level. This size of the grid is used at 7 hectares so that inside each recommended grid, there must be at least 3 to 5 hectare of an area which can be planted. A total number of 795,826 grids were created.

### *2.2. Data and variables*

Due to incomplete and biased field survey data, we used satellite data coupled with GIS for ML training. Open access spatial datasets related to social, biophysical and ecological factors that associate with the long-term success of tree planting were used to create various features used in the recommendation engine. These features were identified based on theory and expert knowledge that determine the availability of blank areas as well as suitability for the long-term successful growth of trees (Supplementary Material Table S1, S2 and exclusion rules given below). We used guidance from the Food and Agriculture Organization and traditional forestry knowledge from forestry field staff to inform our recommendation system (Booth & Saunders, 1985). Trees require adequate temperature, moisture, nutrients, aeration, appropriate radiation, and rooting environment as well as absence of conditions such as adverse soil, climatic and other conditions such as attack by pests and diseases (FAO, 1984). Critical value ranges for various variables used in the analysis differ across various landscapes due to variation in biophysical and socio-economic context, therefore it is not feasible to give specific grid-level details for each variable (FAO, 1984). We use a single value for each variable for all the grids to facilitate the analysis. We store these variables in the PostgreSQL database and then use the PostGIS extension to create various features for each grid using Python. The spatial data was geoprocessed in QGIS and ArcMap.

Variables included in the recommendation engine include forest cover data from Forest Survey of India (FSI), which has time-series cover data on tree density classes. These classes include: open forest (10-40% canopy density), moderately dense forest (40-70%) and very dense forest (>70%), non-forest, scrub, and water. The prior extent of vegetation is a critical factor that determines the extent and possibility of growing trees in any landscape. Open forest patches are potential for growing trees provided they are not natural blanks or rocky lands or other biophysical limits that constraint the successful growth of trees in any landscape. Other important factors that determine the growth of trees include elevation, slope, and aspect of any plantation. Elevation beyond the tree line (~3800 m) and the presence of the southern slope prevent any productive tree growth. We use soil depth and soil carbon (organic and inorganic) as critical determinants that guide whether trees will grow and establish over the long-term. Higher soil depth and soil carbon facilitate the successful establishment of plantations due to higher availability of soil nutrients, soil moisture and humus.

We used guardrails in the forms of exclusion rules, which are based on traditional forestry knowledge to identify blank patches for growing trees in the state. We exclude areas that are covered with grasslands, alpine or sub-alpine pastures, trans-Himalayan zone, natural blanks, snow, agriculture, high resource use, roads, and highways from the potential areas available for tree planting. Natural blanks are areas that are devoid of any vegetation for a long time due to critical limits to tree growth imposed by local environmental and biophysical limits. We identify forest patches (grids) falling inside such natural blank areas by using a set of the expert or traditional forestry rules based on changes in tree density between 2001 and 2019. We also used the number of villages in the vicinity of the assessed forest grid (forest patch) as indicative of higher resource use, which can restrict the success of tree planting over the long-term.

To store and query spatial datasets, we used the PostgreSQL database and its extension called PostGIS, which adds support for geographic objects to the PostgreSQL open-source object-relational database. We created different features based on our spatial queries, which were run on PostGIS using python.

*2.3 Machine learning*

We used ML to predict the likelihood of tree cover loss in each grid (1 = greater than 50% mortality; 0 = less than 50% mortality) falling within a forest compartment using XGBoost, a decision-tree ensemble-based method. We divided the data into 80% training and 20% test datasets for use in the ML model. Our XGBoost model has 0.63 precision and 0.57 recall while predicting the increase in the tree cover in a forest compartment.

Given the heterogeneity of features and the fact that some are categorical, we focused on decision-tree based methods. In this analysis, we had tried two models: Random Forest and XGBoost under standard parameters. Results were similar for both models. Random Forest had 0.63 precision and 0.55 recall for predicted increase in forest cover loss. We chose XGBoost because it has better recall for predicting forest loss compared to Random Forest. In our case, we believe recall is more valuable compared to precision as missing a true positive (tree cover loss) may lead to serious ramifications for biodiversity and forest cover in the area. Moreover, given the heterogeneous trends noticed for forest loss or gain in study region(Rana & Miller, 2019), Random Forest model may under or over predict observations outside of the time frames in our training data depending upon the nature of these trends.

A compartment is the smallest forest management unit. We use 16674 compartments in this analysis. The outcome variable is a dummy variable indicating the increase or decrease in the tree cover loss from 2003 and 2015. ML-based analysis at the compartment level is done to incorporate deforestation (tree cover loss) dynamics and associated factors and processes operating at the landscape level. Incorporation of the deforestation probabilities in the tree planting recommendation engine along with expert-based rules makes the engine more powerful in predicting site suitability classes for growing trees.

In cases where there are multiple grids within a compartment, we consider all these grids to have an increase (or decrease) in tree cover loss if we found an increase (or decrease) in tree cover loss in that compartment. In cases where a single grid is present in multiple compartments, we used the prediction results for the compartment with the largest coverage for that grid.

*2.4. Algorithm fusion and site suitability classification*

Finally, we used scoring rules for each grid in terms of its suitability to grow trees in terms of the presence of blank areas for planting and the site suitability of a site to support tree growth.

Let $X_1$ to $X_n$ be the forest grids (parcels) of size 7.0225 hectares created for the entire state of Himachal Pradesh. Let $S_1$ to $S_n$ be the site suitability scores for each of the forest grid based on the expert/theory rules. The rubric for the scoring grid based on expert/theory is as given in Table S2.

Let $M_1$ to $M_n$ be the site suitability scores for each of the forest grid based on the machine learning model. The final score for each grid is weighted with 90% weight is given to score yielded by theory/expert-based rules whereas 10% weight is given to ML-driven score, *$X_i = 0.9S_i + 0.1M_i$*.

Our choices of weights for expert/theory and ML-driven scores are influenced by large-scale epistemic uncertainties (unknown unknowns) in data-driven assessment of suitability of forestry landscapes for growing trees. That is, available data does not capture many aspects of the socio-ecological and biophysical factors that influence site suitability, but centuries of human experience as distilled into expert Indian Forest Service knowledge do. On the other hand, human knowledge is unable to take large compartment scale deforestation dynamics into account. As such, traditional knowledge and data-driven ML approaches are complementary.

In order to determine the weighting between traditional knowledge rules and the ML algorithm (our final obtained result was 90% for traditional knowledge and 10% for ML), it would have been best to do formal parameter tuning to measure performance criteria like error probability, however, ground truth was lacking in order to do this, as one actually has to plant trees and see how they grow to do this. Instead in our informal parameter tuning approach, we still varied the weighting parameter (among all possible 10% jumps) and instead measured the proportions of area classified to different land suitability categories. This distribution of total forestland among different categories [largely unsuitable, low suitability, medium suitability and high suitability] was then matched to the distributional assessments previously developed by the Indian Forest Service for the state of Himachal Pradesh (Table 1). Through this distributional matching approach, we found that the combination of 90% weight to expert rules and 10% weight to machine learning yielded the best results, in terms of matching to the limits of tree planting growth in Himachal Pradesh, as per the Indian Forest Service.

Table 1: Percentage of suitability among included grids out of the total geographical area of the state (Highly Suitable > 70 score, Medium Suitable > 40 score and <=70 score, Low suitability >0 and <=40 score, Largely Unsuitable=0 score)

| Proportion of rule/expert-based weightage | Proportion of ML weightage | Largely Unsuitable (%) | Low Suitability (%) | Medium Suitability (%) | High Suitability (%) |
|---|---|---|---|---|---|
| 100 | 0 | 68.22 | 18.48 | 11.92 | 1.37 |
| 90 | 10 | 68.46 | 15.71 | 14.15 | 1.68 |
| 80 | 20 | 68.46 | 14.31 | 16.75 | 0.48 |
| 70 | 30 | 68.46 | 13.06 | 18.29 | 0.19 |
| 60 | 40 | 68.46 | 11.48 | 19.97 | 0.15 |
| 50 | 50 | 68.46 | 9.37 | 22.05 | 0.12 |
| 40 | 60 | 68.46 | 6.06 | 25.36 | 0.12 |
| 30 | 70 | 68.46 | 3.57 | 27.85 | 0.12 |
| 20 | 80 | 68.46 | 3.57 | 27.85 | 0.12 |
| 10 | 90 | 68.46 | 3.57 | 27.86 | 0.11 |
| 0 | 100 | 68.46 | 3.57 | 27.57 | 0.40 |

The parameter tuning tried to match total tree growth potential in the state as estimated by the forest service. Out of the total recorded forest area in the state (37,948 $Km^2$), nearly 16,376 $Km^2$ is under alpine pastures, barren lands and perpetual snow. This leaves around 21,572 $Km^2$ area where there exists some kind of tree or forest cover (Himachal Forest Statistics, 2019). Out of this tree growth potential estimate, 3113 $Km^2$ is under very dense forest and 315 $Km^2$ is under scrub, bushes and other perennial shrubs (Forest Survey of India, 2019). Both of these categories have been excluded from the site suitability estimation, leaving a potential tree growth area of around 18144 $Km^2$. Our site suitability application estimated a total tree growth potential area of around 17559 $Km^2$. We attribute the difference (585 Km2) to methodological limitations in tree growth estimations by forest service, which fails to capture few scattered trees existing on largely unsuitable landscapes, which are not fit for any potential tree growth in largely unsuitable category (Forest Survey of India, 2019).

Based on these scores, we then grouped specific sets of grids in 4 classes: largely unsuitable, low suitability, medium suitability and high suitability. Any forest grid with more than 70% score is categorized as high suitable, with >40 and <=70% score is categorized as medium suitable, >0 and <=40 as low suitable for planting tree plantations. Whereas, those grids with zero values are categorized as largely unsuitable. The screenshot of the mobile app (ePSA) is given in Fig. 2 showing different site suitability classes.

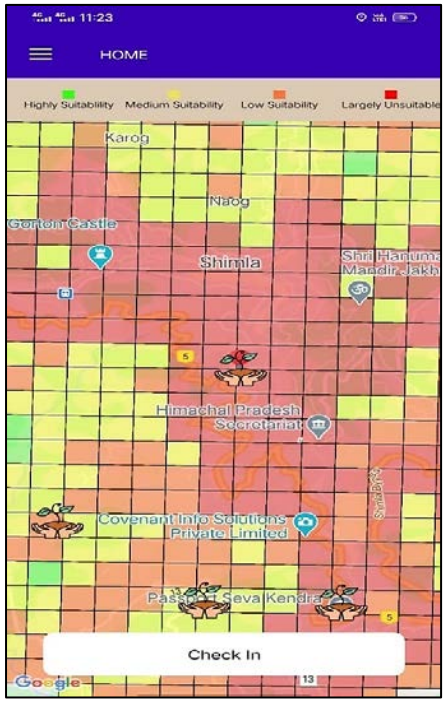

Fig. 2. ePSA (mobile app) showing site suitability classes

## 3. Results

### *3.1. Recommendations for Himachal Pradesh*

Our results show that 68.46% (38,114 km$^2$) of the total area (55,661 km$^2$) in the state of Himachal Pradesh is highly unsuitable for growing trees and the remaining 31.54% area (17,548 km$^2$) is suitable to grow trees The recommendation system prescribes a site suitability score, ranging from 0 to 100, for each forest grid/parcel. Based on this score, we divide the entire area available in the state into four categories to assist tree-planting decisions of forest rangers. These categories are adapted from (FAO, 1984).

- **Largely unsuitable**: Almost inappropriate areas for long-term establishment of plantations due to severe soil and other site quality constraints, and higher plausible biotic pressure
- **Low Suitability**: Areas with the lowest probability of achieving established plantations due to inadequate soil and other site quality conditions, low availability of open areas for growing trees and plausible biotic pressure
- **Medium Suitability**: Areas with an above-average probability of achieving established plantations due to moderate availability of open areas for growing trees and modest soil, other site quality conditions, and biotic pressure
- **High Suitability**: Areas with a high probability of achieving established plantations due to largely supportive soil and other site quality conditions

Out of available 31.54% of the geographical area for growing trees, we found 15.70% (8740.66 km$^2$) has low suitability, 14.14% (7872.16 km$^2$) has medium suitability and 1.68% (935.31 Km$^2$) has high suitability for tree planting in the state of Himachal Pradesh.

The largely unsuitable category includes areas where there is a high percentage of Non-Forest land (84.8%) with very low proportion under OF (Open Forest), MDF (Moderately Dense Forest) and VDF (Very Dense Forest) (Table 2). These areas are also located at a very high elevation (3562 m, average) with about 412,868 grids with villages falling at 1 km or less from these areas, indicating the extent of the pressure of the local population on forest use. The very high percentage under Non-Forest land indicates the presence of rocky, snow-bound, alpine and sub-alpine pastures, croplands and water reservoirs, which restrict any productive tree planting activity in the area.

Table 2: Descriptive on plantation site suitability classification in the state of Himachal Pradesh

| Suitability | Mean OF% | Mean MDF% | Mean VDF% | Mean NF% | Mean elevation in meters | Number of grids with village at 1 km or less |
|---|---|---|---|---|---|---|
| Largely Unsuitable | 4.87% | 4.36% | 2.39% | 84.80% | 3562 | 412868 |
| Low Suitability | 7.00% | 12.37% | 20.30% | 58.42% | 1651 | 332166 |
| Medium Suitability | 25.38% | 53.00% | 5.12% | 5.12% | 1541 | 275110 |
| High Suitability | 76.56% | 22.21% | 0.15% | 0.15% | 1293 | 29158 |

On the other hand, areas with low suitability for tree planting have 58.42% area under Non-Forest, 20.3% under VDF, 12.37% under MDF and 7% under OF (Table 1). These areas are located on an average elevation of around 1651 m with 332,166 grids with villages located at 1 km or less. Medium suitable areas have a high percentage under MDF (53%), followed by OF (23.38%) and then VDF (5.12%). The percentage under NF is low (5.12%), which indicates the presence of areas for planting trees. Such areas are located an average elevation of about 1541 m with 275,110 grids with villages are falling at 1 km or less from these areas.

Lastly, areas with high suitability have a very high percentage of OF (76.56%), followed by MDF (22.2%) and VDF (0.15%). Only 0.15% area is falling under NF, which indicates the high suitability of the area for planting trees (Table 1). These areas are located at a mean elevation of about 1293 m and only 29,158 grids with villages are located at 1 km or less from these areas.

*3.2 Tree planting in Monsoon 2020*
Throughout the state of Himachal Pradesh, we found 25.4% of the total proposed tree plantations in largely unsuitable category ($n = 393$), 40.6% in areas with low suitability predictions ($n = 627$), 33.2% in areas with medium suitability ($n = 513$) and only 0.9% in areas with high suitability ($n = 14$) (Fig. 3).

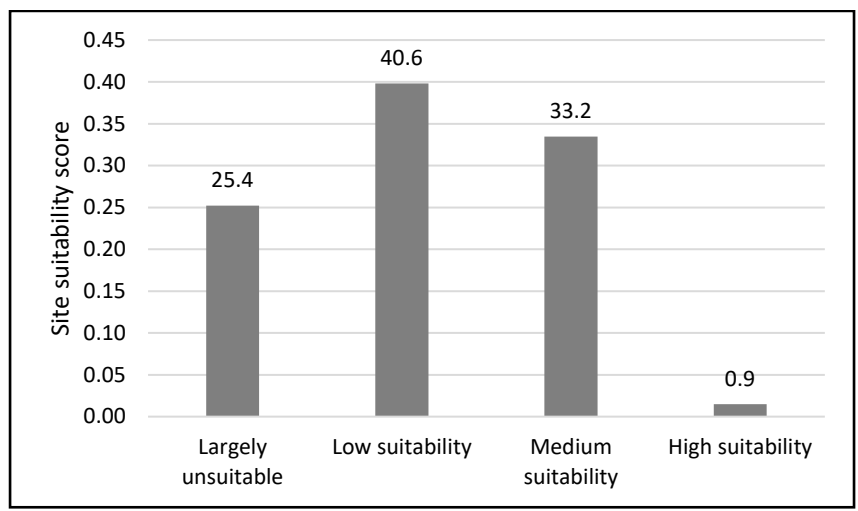

Fig. 3. Site suitability scores for each of the four site suitability classes as per the recommendation engine

The descriptive for the plantation sites (*n* = 1546), planted by the forest department, during monsoon 2020 is given in Table 3. The largely unsuitable class has a very high percentage of area under Non-Forest (NF) (70.1%), followed by low suitability (47.4%) and medium suitability classes (20.4%), with high suitability class has a very low area under Non-Forest (4.9%). Both largely unsuitable and low suitability areas are located at a higher elevation, have a low percentage of area under forest cover and have a considerable number of villages near their vicinity. On the other hand, medium and high suitability areas have a very high percentage of Open and Moderately Dense Forests and are located at lower elevations (Table 3).

Table 3: Distribution of planting sites (*n* = 1546) in different suitability classes as per the recommendation engine and their attributes (Fig. 4)

| Suitability classes | Number of proposed tree plantation sites | Open Forests (OF), mean, (%) | Mod. Dense Forests (MDF), mean, (%) | Very Dense Forests (VDF), mean, (%) | Non-Forest (NF), mean, (%) | Elevation (meters, average) | Number of grids with a village at 1 km or less |
|---|---|---|---|---|---|---|---|
| Largely unsuitable | 392 | 13.6 | 10.9 | 3.7 | 70.1 | 1825 | 304 |
| Low suitability | 627 | 13.9 | 20.7 | 17.2 | 47.4 | 1553 | 432 |
| Medium suitability | 513 | 24.1 | 49.5 | 5.6 | 20.4 | 1359 | 338 |
| High suitability | 14 | 58.1 | 35.5 | 1.4 | 4.9 | 1291 | 7 |

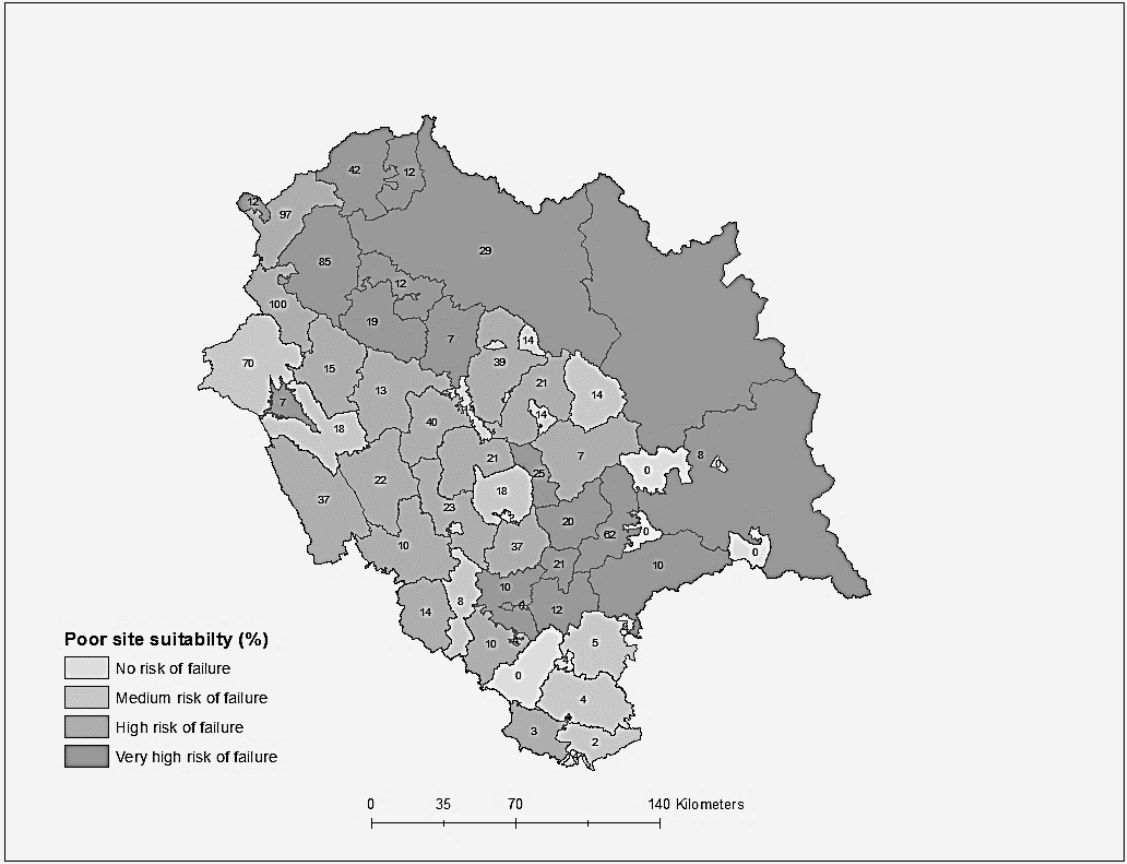

Fig. 4. The number of sites selected in largely unsuitable and low suitable classes in 43 forest divisions during Monsoon 2020. The divisions with a number represent areas where either there is no data is available or where winter planting is normally done. The darker the shade, the higher is the number of sites being planted in areas with a higher predicted risk of failure due to poor site suitability values as per ePSA.

### 3.3. Feedback from field officials: Lessons learned

In this section, we present feedback from field officials and the lessons learned after the deployment of the plantation site recommendation engine. This is from a random sample of 30 forest officials out of 201 users (15%) about the mobile app across the state of Himachal Pradesh. A phone survey was conducted in September 2020 and forest officials were asked to explain a) main features that they liked about the application and whether machine recommendations matched field realities, b) the main weaknesses of the application while it on the ground and, c) possible improvements to increase the adoption of the application in future. Only 4 officials disagree with the recommendations of the ePSA whereas 26 officials fully agreed with the suggestions by the recommendation engine. This means 86.67% of the officials were satisfied with recommendations, showing higher reliability of the recommendation engine for growing trees.

Care was taken not to bias the user study findings due to the position of the first author, and we believe there is no such bias in the responses of the forest officials with respect to their feedback about ePSA. While interviewing, the identity of the first author was not disclosed to forest officials. Moreover, no interviewed official works directly under the first author.

Responses can be categorized based on the level of satisfaction while using ePSA, as follows.

a) **Satisfied**: 26 forest officials were fully satisfied with the recommendations made by ePSA and found them to adequately represent the field site conditions for growing trees. According to the officials, ePSA could suggest appropriate locations for planting trees, assist in holistic management of plantation activities and provide direct data and feedback to higher officials. Other benefits included loading of photos at one time, ease in getting latitude and longitude of the location automatically and learning about the importance of site conditions such as slope, aspect, and other factors that lead to higher suitability of a site for growing trees.

   Some comments from the forest officials about the working of the app include:

   "ePSA is really good. The accuracy of location is good. It is very good for selecting site and also, option to click photos covering North, East, West and South is good. Site suitability predictions are also as per the field conditions and it is accurately showing where density of tree cover is more. It also helps in identifying appropriate blank patches for growing trees." (Forest Guard, Dhanni)

   "We get full information about the plantation site where plantations is being planted. We are also sending full details along with photos through this application. The site suitability recommendations are as per the field site conditions. According to me, there is no shortcoming in this application. Moreover, it allows us to send to higher officials whether we have area available on ground for plantations or not, which will help in better plantation planning in future" (Forest Guard, Dhamla Beat)

   "The app is good. It shows the shortcomings in the site location for growing trees. I agree with the recommendation with the app. The process is good and it captures information about fire incidents, grazing and labor availability for growing trees." (Forest Guard, Badukhar Beat).

   "App is good as it can help higher offices monitor the progress of plantations. App is showing low suitability at many places. By the time the app was launched, we had already completed plantations. Plantation on deodar is good on the other side but app is continuously suggesting non-suitability of sites. We came to know about this app later on and could not change the site for planting trees". (Forest Guard, Bijlidhar).

   "We can inform the higher officers about our work. We also do not pay attention to site factors while planting trees. Normally, forest guard alone takes decision. Now, app will help him take decisions on ground." (Forest Guard, Mangwal).

   "App is good. Recommendations are good. In hilly terrain, sloppy areas are more. We do not get full blank areas on the ground. In google, we see blank areas from top but on ground, we do get areas for planting trees on ground. The area calculated through GPS in sloppy area is shown less but in actuality the calculated area is more". (Forest Guard, Shalogani).

   "I have tried at multiple locations. The recommendations were matching with field realities at the sites that I visited. It seems the recommendations were based tree density and soil conditions. I found medium suitable in good soil conditions and less suitable in rocky patches. Overall, the app was matching the field conditions." (Divisional Forest Officer, Sangrah).

"The entire details of the plantation is available and criteria is good. There will be benefit in the ground to use it as it has all the criteria given for choosing a right site for planting trees." (Forest Guard, Bijhari Beat).

b) **Unsatisfied**: Some of the forest officials found ePSA as not user-friendly and mentioned the crashing of application while loading pictures. Moreover, they raised questions about the predictions made by ePSA compared to their own judgement about the site quality of the planted area. The comments by these forest officials are given as:

"We were using the app, but we were unable to upload photos in the said application. The app is not running properly. I do not agree to the suggestions offered by the app as there were some good patches to grow trees though there were rocky areas in the middle of the plantation site. But, the app totally rejected the site. There is about 90% area where tree planting can take place as per my view. There was only 10% rocky stones towards the side of the hill. I believe that the plantation in this site will definitely succeed. Moreover, app is crashing while uploading photos." (Forest Guard, Sangholi Forest Beat).

"The app is suggesting non-suitable areas for planting through the Nurpur division. The app is suggesting high suitable to pockets where there are already good tree cover and not suitable for areas where there are not good tree cover. I went to sites where we have carried out plantations but still the app is suggesting those sites as not suitable. And these sites are shown as non-suitable despite good tree survival on these sites. Most of the land in the Nurpur division is rocky but those rocks are near the valleys but the soil on top is really fertile. We have not checked all sites inside Nurpur. I have checked an area where there were large chir pine plantation but now was burnt. But, the app was suggesting this area as not suitable. The density is even less than moderately dense class. (Forest Guard, Aund Beat)

"Accessing ePSA is problematic. Offline loading of photos is required. Recommendations are not so suitable to my area. My area is good as it used to be Chir Pine forest in the past but now the app is ranking it as non-suitable". (Forest Guard, Gharanu Beat).

"The app was recommending the area as non-suitable. The area is fit and lantana is removed for planting trees. We took 8 hectares and we then used 5 hectares as final area for planting. We do not have suitable areas for planting trees and most of the areas are rocky. We already selected area before the launching the app" (Divisional Forest Officer, Kunihar for Danoghat site).

Officials suggested improving the user experience through removal of bugs such as crashing of app while loading photos, offline loading of photos, offline use of ePSA, updating information, providing user-friendly access to app and showing large area for view while locating suitable site for planting trees. Some suggest linking Compartment History File data with ePSA data to understand the actual site condition and recommendation based on this data would be useful for any prediction regarding site suitability. Adding such features in future would further increase the practicality and importance of ePSA in suggesting suitable sites for planting trees. Some officials suggested use of app about two months before actual plantations to fully explore the plantation landscape.

## 4. Discussion and conclusion

Our analysis highlights the role of algorithm fusion in designing a plantation site recommendation engine based on the coupling of theory and expert rules with machine learning. Due to the lack of data and social-ecological complexity in forest systems, we relied on both satellite data and field expert knowledge to create the ePSA, recommendation system. After database creation and feature engineering based on expert knowledge, we used machine learning to capture landscape-level deforestation outcomes in a landscape to guide our recommendations. Our paper suggests the utility of the algorithm fusion approach from the AI safety literature in dealing with complex social-ecological problems such as selecting suitable sites for planting trees (Kshetry & Varshney, 2019).

Our results indicate that there is a very high percentage of tree planting happening in places where there is a higher likelihood of plantation failure and wastage. Our initial evidence indicates that most of the officials found ePSA very helpful in choosing the right sites for planting trees, getting a broader view of the plantation site, communicating feedback to higher ups, and assisting in holistic plantation management. In 26 out of 30 queried locations, forest officials found the recommendations of the app very useful for selecting the right site for planting trees. Although there are some changes required in the operationalization of the app, ePSA proved to be very useful for field staff on the ground. However, full operationalization requires early selection of plantation sites much before the plantation season to give ample time to forest officials to scan the area with app to choose the right location for planting trees.

Due to COVID-19 restrictions, there was delay in launching the application, so only 201 out of about 1000 final users could test it on the ground. We do not believe there is any systematic bias in this subset. In our user study to assess the usefulness of ePSA, we surveyed 30 out of these 201 users (15%) via a random sample. Though, we admit our constraints in obtaining empirical data on the quality of suitability prediction generated and the fact we used the expert feedback to confirm ePSA quality. We admit that it would be fruitful to revisit his topic in the future to evaluate the progress of recommendations on the ground and the tree growth results. Note that tree growth itself does take some years.

Notably, our user study (which is carried out using standard methodology for such research in the human-computer interaction (HCI) literature and not just unstructured anecdotes) indicates that it was not just the information provision through ePSA that was useful, but also the AI-based assessments. In particular, forest officials in 11 sites shifted to new locations for planting trees based on the AI assessment. This is a significant fraction, given the level of professional prestige that Forest Service officers enjoy in India, and given the well-known phenomenon that people in high-prestige professions (such as doctors, judges, and astronauts) are most likely to assert their own judgements ahead of AI (see e.g. Neil Armstrong landing on the moon by joystick rather than autopilot, as described in (Mindell, 2011)).

We also noticed that some of the field officials avoided using ePSA and continued to plant trees in unsuitable locations in many cases. This trend may be due to several reasons. First, as our results indicate that only a small percentage of area in the state of Himachal Pradesh has high suitability (1.68%), where trees (1100 trees per hectare) can be planted. This means there is an overall absence of large patches for forests for planting trees across the state and, due to the target-driven approach of tree-planting programs, forest rangers have no other alternative other than to continue to select pockets where either tree planting is unnecessary due to high existing forest cover or due to poor site factors such as slope, elevation, soil quality, and other features (Fleischman, 2014; Saxena, 1997).

Second, there are no provisions for planning site-specific tree planting as per existing vegetation, resource use, and site factors. Tree planting programs and norms are imposed by higher-ups uniformly across the state without carrying out a prior site assessment and diagnostic to design such programs as per the site requirements (Fleischman, 2014). In many such places, a higher focus on assisted natural regeneration (ANR) rather than planting trees can be more effective and efficient to protect local biodiversity and wastage of financial resources (Crouzeilles et al., 2020; Duguma et al., 2020).

Third, governments and donors usually evaluate the performance of forest rangers and other officials based on accountability systems that emphasize targets for the number of trees or areas under plantation (Fleischman, 2014). Such tree planting works are visible to external agencies for monitoring and evaluation and therefore, are promoted on large scale with forest rangers and other forest officials competing with one another in terms of the number of trees and area planted even if suitable areas for growing trees are not available on the ground. Finally, poor institutional incentives, rewards and punishment systems in the forest department encourage large-scale tree planting at the costs of local livelihoods, biodiversity, and tree cover. The vested economic interests of the forest rangers motivate them to prefer tree planting and soil conservation works (e.g. check dams) over assisted natural regeneration, silviculture operations, effective fire and grazing management, or other forest improvement activities (Fleischman, 2014).

Our approach has some limitations. First, satellite data may miss some important social-ecological dynamics due to the choice of specific spatial and temporal resolutions, incompatibility in various spatial layers, and radiometric corrections. For example, a stunted Chir Pine vegetation due to local biophysical constraints may appear to be a patch of scrubland or areas with scarce trees, which the algorithm may suggest as a potential patch for plantation. Second, the App does not prescribe the number and type of tree species be planted in a particular site. Third, regular diversion of forestland to non-forestry uses such as constructing roads, dams, buildings etc., salvage tree removal by state forest corporation, and forest loss due to wildfires, floods and diseases require continual updating of the App to make it more dynamic and suited to local field realities. Machine learning systems, which rely on large amounts of data may have led to a flexible, adaptable and dynamic recommendation system. However, in our case, we could not develop a fully machine-learning-based system due to the pervasive epistemic uncertainty in tree plantation domain, absence of appropriate fine-resolution data and field survey assessments, social-ecological complexity of tree planting decisions and incompatibility of various satellite-based spatial layers. Finally, ePSA needs to be further tested on the ground to evaluate its performance under diverse settings and to improve its performance.

Despite all the above limitations, ePSA provides a promising algorithmic fusion-based recommendation system to suggest tree planting, which can assist forest rangers to identify forest patches for tree planting as per availability of blank areas and site suitability classification. The initial field assessments of ePSA show the relevance of the recommendation system to inform appropriate site selection decisions, which can save a lot of financial resources to achieve effective carbon sequestration while protecting local biodiversity and livelihoods. Importantly, such recommendations are practical and highly relevant to forest rangers as these prescriptions are based on evaluating a particular forest patch (~7 hectares) for exploring a patch of 3 to 5 hectares for tree planting, which is an average size for planting trees in the state of Himachal Pradesh.

Shifting the focus from extensive tree-planting to protecting forest ecosystem goods and services, supporting forest-based livelihoods, prioritizing assisted natural regeneration and promoting broadleaved species seems the most appropriate future direction for the state of Himachal Pradesh. In this context, ePSA can play an important role in prioritizing suitable areas for growing trees to minimizing waste, maximizing co-benefits, and evaluating such tree planting activities. The practical utility of ePSA indicates its promising use in largely similar contexts of other developing countries in the world, which are also witnessing similar large-scale industrial or small-scale public tree planting programs, and can help avoid massive waste on tree planting programs with little contribution to carbon storage and local livelihoods.

**Acknowledgements**

We thank Himachal Pradesh Forest Department for sharing the data. We are grateful to FIRST, IIT-Kanpur, India for providing help in processing and analyzing the data and assisting in the development of ePSA. We also thank all the forest officials of Himachal Pradesh Forest Department, who have contributed in the development of ePSA through their knowledge, usage and feedback.

Kshetry, N., & Varshney, L. R. (2019). Safety in the face of unknown unknowns: Algorithm fusion in data-driven engineering systems. *ICASSP 2019-2019 IEEE International Conference on Acoustics, Speech and Signal Processing (ICASSP)*, 8162–8166.

Lahssini, S., Lahlaoi, H., Alaoui, H. M., Bagaram, M., & Ponette, Q. (2015). Predicting cork oak suitability in Maamora forest using random forest algorithm. *Journal of Geographic Information System*, *7*(02), 202.

*Livestock Census | Department of Animal Husbandry & Dairying*. (n.d.). Retrieved February 29, 2020, from http://www.dahd.nic.in/documents/statistics/livestock-census

Malkamäki, A., D'Amato, D., Hogarth, N. J., Kanninen, M., Pirard, R., Toppinen, A., & Zhou, W. (2018). A systematic review of the socio-economic impacts of large-scale tree plantations, worldwide. *Global Environmental Change*, *53*, 90–103. https://doi.org/10.1016/j.gloenvcha.2018.09.001

Mallick, A., Dwivedi, C., Kailkhura, B., Joshi, G., & Han, T. (2020). Probabilistic Neighbourhood Component Analysis: Sample Efficient Uncertainty Estimation in Deep Learning. *ArXiv Preprint ArXiv:2007.10800*.

Mindell, D. A. (2011). *Digital Apollo: Human and machine in spaceflight*. Mit Press.

Mueller, S. T., Hoffman, R. R., Clancey, W., Emrey, A., & Klein, G. (2019). Explanation in Human-AI Systems: A Literature Meta-Review, Synopsis of Key Ideas and Publications, and Bibliography for Explainable AI. *ArXiv Preprint ArXiv:1902.01876*.

National Centers for Environmental Information, National Oceanic and Atmospheric Administration. (n.d.-b). *Version 4 DMSP-OLS Nighttime Lights Time Series*.

Rana, P., & Miller, D. C. (2019). Explaining long-term outcome trajectories in social–ecological systems. *PloS One*, *14*(4), e0215230.

Sabatia, C. O., & Burkhart, H. E. (2014). Predicting site index of plantation loblolly pine from biophysical variables. *Forest Ecology and Management*, *326*, 142–156.

Salganik, M. J. (2019). *Bit by bit: Social research in the digital age*. Princeton University Press.
19

**Supplemental materials**

**Table S1.** Predictor variables and their sources

| Variable | Description | Unit of measurement | Sources of data |
| --- | --- | --- | --- |
| Number of households | Number of households | Total number of HHs in villages that are inside a forest polygon | Census (2001), India, http://censusindia.gov.in/ (*Census of India, 2001*, 2001) |
| Total population | Total population | Total population of the villages that fall inside a forest polygon | Census (2001), India, http://censusindia.gov.in/ (*Census of India, 2001*, 2001) |
| Number of farmers | Number of cultivators (farmers) | Total number of farmers in villages that fall inside a forest polygon | Census (2001), India, http://censusindia.gov.in/ (*Census of India, 2001*, 2001) |
| Number of marginal people (scheduled caste population) | Scheduled caste population | Total number of total SC population in villages that fall inside a forest polygon | Census (2001), India, http://censusindia.gov.in/ (*Census of India, 2001*, 2001) |
| Number of literates | Total number of literates | Total number of literates in villages that fall inside a forest polygon | Census (2001), India, http://censusindia.gov.in/ (*Census of India, 2001*, 2001) |
| Number of unemployed people | Total marginal workers | Total number of marginal workers in villages that fall inside a forest polygon | Census (2001), India, http://censusindia.gov.in/ (*Census of India, 2001*, 2001) |
| Economic activity | 2003–2008, 0.56 km spatial resolution | 1 to 63 (values) (Average for villages that fall inside a forest polygon) | Version 4 DMSP-OLS Nighttime Lights Time Series (n.d.-b) |
| Road density | Road density, | $Km/km^2$ (Average for villages that fall inside a forest polygon) | CIESIN (Data Center in NASA's Earth Observing System Data and Information System (EOSDIS)) (https://sedac.ciesin.columbia.edu/data/sets/browse )(n.d.-a) |
| Number of small landholdings | Number of small land-holdings less than 0.5 ha | Number of smallholdings less than 0.5 ha in Census Tehsils where that forest polygon falls. | Agricultural census (2005), India (*Agricultural Census, 2005*, n.d.) |

| Variable | Description | Unit of measurement | Sources of data |
| --- | --- | --- | --- |
| Grazing density | Number of grazing animals (buffaloes, goats, sheep, cattle)/area of the tehsil in ha | Number/ha (Average for villages that fall inside a forest polygon) | Livestock census (2007), India (*Livestock Census / Department of Animal Husbandry & Dairying*, n.d.) |
| Area of forest polygon | Area of the forest polygon | ha | Forest records, HP Forest Department, India |
| Area under crop acreage | 2000, 30 m resolution | ha | (J. Chen et al., 2015) |
| Area under grass coverage | 2000, 30 m resolution | ha | (J. Chen et al., 2015) |
| Area under bare land acreage | 2000, 30 m resolution | ha | (J. Chen et al., 2015) |
| | | | (Wieder et al., 2014) |
| Soil depth | 2000, reference soil depth, average | cm | (Wieder et al., 2014) |
| Available soil water capacity | 2000, available soil water storage capacity, average | Coded values 1 to 7; 1 = 15 cm water per m of the soil unit, 2 = 12.5 cm, 3 = 10 cm, 4 = 7.5 cm, 5 = 5 cm, 6 = 1.5 cm, 7 = 0 cm. | (Wieder et al., 2014) |
| Topsoil Carbon Content | Topsoil and subsoil carbon content (T_C and S_C) are based on the carbon content of the dominant soil type in each regridded cell rather than a weighted average. | kg C m-2 | (Wieder et al., 2014) |
| Subsoil Carbon Content | | kg C m-2 | (Wieder et al., 2014) |
| Topsoil Organic Carbon | | % weight | (Wieder et al., 2014) |
| Subsoil Organic Carbon | | % weight | (Wieder et al., 2014) |
| PH (Top Soil) | Topsoil pH (in H2O) | -log(H+) | (Wieder et al., 2014) |
| Top Soil Bulk Density | Reference bulk density values are calculated from equations developed by Saxton et al. (1986) that relate to the texture of the soil only. | kg dm-3 | (Wieder et al., 2014) |
| Top Soil Cation Exchange Capacity | Cation exchange capacity of the clay fraction in the topsoil | cmol per kg | (Wieder et al., 2014) |
| Sub Soil Cation Exchange Capacity | Cation exchange capacity of the clay fraction in the subsoil | cmol per kg | (Wieder et al., 2014) |

| Variable | Description | Unit of measurement | Sources of data |
|---|---|---|---|
| Location (altitude) | 2000, 90 m resolution | m | SRTM (Shuttle Radar Topography Mission), 90 m resolution, 2000 (*SRTM 90m Digital Elevation Database v4.1*, 2017) |
| Slope | 2000, 90 m resolution | degree | SRTM (Shuttle Radar Topography Mission), 90 m resolution, 2000 (*SRTM 90m Digital Elevation Database v4.1*, 2017) |
| Baseline forest cover (FC_2003HA) | 2003, 24 m resolution | Forest cover = Open forest + Moderately dense forest + Very dense forest | Forest Survey of India, 2005 (Forest Survey of India, 2019) |
| Number of forest fires | 2003–2008 | Number | NASA, active fire data, MODIS C6 (*FIRMS*, n.d.) |
| Temperature | 2001–2008, 30 km resolution, average | °C | CRU (Climatic Research Unit) TS dataset, version 4.0, gridded dataset of monthly terrestrial surface climate http://www.cru.uea.ac.uk/ (Harris et al., 2014) |
| Precipitation | 2001–2008, 30 km resolution, average | mm | CRU (Climatic Research Unit) TS dataset, version 4.0, gridded dataset of monthly terrestrial surface climate http://www.cru.uea.ac.uk/ (Harris et al., 2014) |
| Land surface temperature | 2001–2008, 5.5 km spatial resolution, average | K | MODIS/Aqua Land Surface Temperature/Emissivity Monthly L3 Global CMG V005 (*Global Change Master Directory (GCMD)*, n.d.) |
| *Outcomes (O)* | | | |
| Tree cover loss | 24 m resolution FC_CHANGE15_03 = FC_2015HA – FC_2003HA | If FC_CHANGE15_03< 0, Tree cover loss = 1, Otherwise = 0 | Forest Survey of India (2005); Forest Survey of India (2017) (Forest Survey of India, 2019) |

**Table S2**: Site suitability scoring rubric based on theory/expert-based rules

| Weightage | Data | Rule | Example |
|---|---|---|---|
| 50% | 2019 MDF % and OF % | Current MDF and OF will contribute to 50 % OF the score | MDF = 40 % and OF = 50% then score will be 0.5*(40+50) i.e. 45 |
| 3% | Slope | Slope <=30, 1.5 point<br>Slope <=50 1 point<br>Slope >50, 0.5 point | Slope = 25 then the score will be 1.5 |
| 3% | Aspect | North Aspect,1.5 point<br>East Aspect, 1 point<br>South Aspect, 0.5 point | Aspect = northern then score is 1.5 |
| 3% | Elevation | Elevation<=1000, 1.2 point<br>Elevation <=2000 0.8 point<br>Elevation <=2500, 0.6 point<br>Elevation >2500 0.4 point | Elevation = 1800 then score is 0.8 |
| 2% | Mean inorganic carbon content | Mean inorganic carbon<=1.5, 0.4 point<br>Mean inorganic carbon<=4.5, 0.6 point<br>Mean inorganic carbon>4.5, 1 point | Mean inorganic carbon = 3.5 then score 1 |
| 2% | Mean organic carbon content | Mean organic carbon<=5, 0.4 point<br>Mean organic carbon<=15, 0.6 point<br>Mean organic carbon>15, 1 point | Mean organic carbon = 20 then score 1.5 |
| 6% | Soil Depth | Soil depth<50cm, 1 point<br>Soil depth<100cm,2 point<br>Soil depth> 100cm,3points | Soil depth = 90 cm then score = 2 points |
| 5% | Nearness OF village from a grid | The closest village is less than 1 km,1 point<br>The closest village is less than 3 km,2 points<br>Closest village in more than 3 km, 3 points | The closest village is 9 km away then score will be 3 points |
| 5% | Forest cover change from 2015 to 2019 | change in MDF > 0 positively scored<br>change in OF > 0 positively scored<br>change in VDF > 0 positively scored<br>change in water>0 negatively scored<br>change in scrubs > 0 negatively scored<br><br>change in OF/MDF > change in VDF/water/scrub positively scored | change in MDF = 40% (0.3-point score will be given)<br><br>change in OF = 10% (0.3-point score will be given)<br><br>change in MDF > change in water (0.3 score point will be given) |
| | | change in NON-FOREST > change in OF/MDF positively scored | |
| | | change in VDF > change in OF/MDF positively scored | |
| | | change in VDF/water/scrub > change in OF/MDF negatively scored | |
| 4% | Forest cover change from 2013 to 2019 | change in MDF > 0 positively scored<br>change in OF > 0 positively scored<br>change in VDF > 0 positively scored<br>change in water>0 negatively scored<br>change in scrubs > 0 negatively scored | change in MDF = 40% (0.26-point score will be given) |
| | | change in OF/MDF > change in VDF/water/scrub positively scored | change in OF = 10% (0.26-point score will be given) |
| | | change in NON-FOREST > change in OF/MDF positively scored | change in MDF > change in water (0.26 score point will be given) |
| | | change in VDF > change in OF/MDF positively scored | |
| | | change in VDF/water/scrub > change in OF/MDF negatively scored | |
| 3% | Forest cover change from 2009 to 2019 | change in MDF > 0 positively scored<br>change in OF > 0 positively scored<br>change in VDF > 0 positively scored<br>change in water>0 negatively scored<br>change in scrubs > 0 negatively scored<br>change in OF/MDF > change in VDF/water/scrub positively scored | change in MDF = 40% (0.2-point score will be given) |
| | | | change in OF = 10% (0.2-point score will be given) |
| | | change in NON-FOREST > change in OF/MDF positively scored | |
| | | change in VDF > change in OF/MDF positively scored | change in MDF > change in water (0.2 score point will be given) |
| | | change in VDF/water/scrub > change in OF/MDF negatively scored | |
| 2% | Forest cover change from 2005 to 2019 | change in MDF > 0 positively scored<br>change in OF > 0 positively scored<br>change in VDF > 0 positively scored<br>change in water>0 negatively scored<br>change in scrubs > 0 negatively scored<br>change in OF/MDF > change in VDF/water/scrub positively scored | change in MDF = 40% (0.13-point score will be given) |
| | | | change in OF = 10% (0.13-point score will be given) |
| | | change in NON-FOREST > change in OF/MDF positively scored | |
| | | change in VDF > change in OF/MDF positively scored | change in MDF > change |

| | | change in VDF/water/scrub > change in OF/MDF negatively scored | in water (0.13 score point will be given) |
|---|---|---|---|
| 1% | Forest cover change from 2003 to 2019 | change in MDF > 0 positively scored<br>change in OF > 0 positively scored<br>change in VDF > 0 positively scored<br>change in water>0 negatively scored<br>change in scrubs > 0 negatively scored<br>change in OF/MDF > change in VDF/water/scrub positively scored | change in MDF = 40% (0.06-point score will be given) |
| | | change in NON-FOREST > change in OF/MDF positively scored | change in OF = 10% (0.06-point score will be given) |
| | | change in VDF > change in OF/MDF positively scored | change in MDF > change in water (0.06 score point will be given) |
| | | change in VDF/water/scrub > change in OF/MDF negatively scored | |
| 1% | Forest cover change from 2001 to 2019 | change in MDF > 0 positively scored<br>change in OF > 0 positively scored<br>change in VDF > 0 positively scored<br>change in water>0 negatively scored<br>change in scrubs > 0 negatively scored<br>change in OF/MDF > change in VDF/water/scrub positively scored change in NON-FOREST > change in OF/MDF positively scored | change in MDF = 40% (0.06-point score will be given) |
| | | | change in OF = 10% (0.06-point score will be given) |
| | | change in VDF > change in OF/MDF positively scored | change in MDF > change in water (0.06 score point will be given) |
| | | change in VDF/water/scrub > change in OF/MDF negatively scored | |